\documentstyle[12pt]{article}
\makeatletter
\@addtoreset{equation}{section}
\makeatother

\newcommand{\half}{\mbox{\small $\frac{1}{2}$}}
\newcommand{\PL}{{Phys.\ Lett.\ }}
\newcommand{\PR}{{Phys.\ Rev.\ }}

\newcommand{\PRL}{{Phys.\ Rev.\ Lett.\ }}

\begin{document}
\thispagestyle{empty}
\begin{flushright}
BI-TP 94/36\\
UCT-TP 213/94\\
hep-th/9408249
\end{flushright}
\vspace*{31mm}
\begin{center}
{\Large \bf
 Gluon Decay as a Mechanism for\hfil\\
 Strangeness Production in a Quark-Gluon Plasma.
}\\
\vspace*{25mm}
{\large  N.~Bili\'c$^{1,2}$, ~J.~Cleymans$^1$,
I.~Dadi\'c$^{2,3}$ and  D.~Hislop$^1$}
\\[24pt]
$^1$Department of Physics,
University of Cape Town, Rondebosch, South Africa  \\
$^2$ Rudjer Bo\v{s}kovi\'{c} Institute,  Zagreb, Croatia\\
$^3$ Fakult\"at f\"ur Physik, Universit\"at Bielefeld , Germany \\
\vspace*{25mm}
\end{center}
\begin{abstract}
A calculation of
thermal gluon decay shows
that this process contributes significantly
to strangeness production in a quark-gluon plasma.
Our analysis
does not support recent claims that this is the
dominant process.
In our calculations we
take into account the resummed form of the
transverse and longitudinal parts of the gluon propagator
following the Braaten-Pisarski method.
Our results
are subject to the  uncertainty concerning
the estimate of the damping rate  entering the effective gluon
propagator.
\end{abstract}
\newpage
\section{Introduction}
A possible signal for quark-gluon plasma formation
in heavy-ion collisions is
the enhancement of the production of
strange particles.
The original proposal by Rafelski and M\"uller
 \cite{rafelski1}
was followed by extensive discussion in the
literature \cite{biro1,rafelski2,matsui,biro2,altherr1,altherr2}.
In this context thermal gluon decay has been recently
discussed\cite{biro2,altherr1,altherr2}.
It has been claimed recently that the process
$g\rightarrow\bar{q} q$ dominates for a wide range
of quark masses \cite{altherr1,altherr2}.
Normally, the gluon cannot decay into a strange
quark-antiquark pair because its thermal mass is too low.
Even for the optimistic case where on takes the coupling
constant $g=2$ in a plasma with two
massless quarks, the gluon mass is given to lowest order
in perturbation theory by
\begin{equation}
m_g={2\over 3}gT.
\end{equation}
For a temperature of $T$= 200 MeV this gives $m_g = 267$ MeV
which is  below the threshold for the production of strange
quarks.
The important observation  by Altherr and Seibert is that
in addition to
 acquiring a thermal mass,
 gluons
also acquire a  width, determined by the large
damping rate, of the order $g^2 T$\cite{pisarski}.
%
%
%This can be seen very clearly in Fig. (1) where the
%transverse part of the gluon propagator
%is plotted versus the invariant mass of the gluon. The peak of the
%distribution is clearly below the threshold for $s\bar{s}$ production
%but there exists a long tail above this threshold.
%
%
%
%
 This is the reason why thermal gluon decay into
a heavy quark-antiquark pair is allowed, even though the gluon mass is
below the threshold for strange pair production.

In this paper we present a systematic re-evaluation
of the production rate of massive quarks
in a quark-gluon plasma due to the processes
   of quark-antiquark annihilation,
gluon fusion and thermal gluon decay
in the spirit of Altherr and Seibert.
 Since the production rate depends strongly on the
 damping rate
we take a more conservative approach in its estimation.
Our main point is that even with the parameters chosen in Ref.
\cite{altherr1,altherr2}
 we cannot support the claim that  gluon
decay is the dominant mechanism for strange quark production
in a quark-gluon plasma.
To the best of our knowledge, the gluon
fusion mechanism, originally proposed by Rafelski
and M\"uller\cite{rafelski1},
remains the leading process.

The paper is  organized  as follows. In section 2 we
briefly review
the properties of thermal-gluon propagators and the damping
rate. In section 3 we calculate the production rates.
In section 4 we present results and concluding remarks.
\section{Gluon Propagator and  the Damping rate}
The effective gluon propagator
at finite temperature in the Feynman gauge is
given by
\cite{weldon1}
\begin{equation}\label{eq38}
i D_{\mu\nu}^{ab}(q_0,q)=
-i \delta^{ab}
[P_{\mu\nu}^T \Delta_T(q_0,q)+
P_{\mu\nu}^L \Delta_L(q_0,q)] ,
\end{equation}
where
      $P_{\mu\nu}^T$
and   $P_{\mu\nu}^L$
are transverse and longitudinal projectors respectively, and
\begin{equation}\label{eq38a}
\Delta_{T,L}(q_0,q)=\frac{1}{Q^2-\Pi_{T,L}(q_0,q)}.
\end{equation}
where $Q^2\equiv q_0^2-q^2$.
The real transverse and longitudinal
parts of the gluon self energy in the high
temperature limit are respectively given  by
\begin{equation}\label{eq39}
\mbox{Re} \Pi_T(q_0,q)=
{3\over 2}m_g^2\left[ {q_0^2\over q^2}+\left( 1 - {q_0^2\over q^2}
\right) {q_0\over 2 q}\ln {q_0 +q\over q_0 -q} \right],
\end{equation}
and
\begin{equation}\label{eq310}
\mbox{Re} \Pi_L (q_0,q)={3\over 2} m_g^2
\left( 1 - {q_0^2\over q^2} \right)
 \left[2- {q_0\over q}
\ln {q_0 + q\over q_0 -q} \right] .
\end{equation}
The positions of the poles
 in the propagator (\ref{eq38a})
 are determined by the
dispersion relations
\begin{equation}\label{eq311}
q_0^2=q^2+
\Pi_{T,L}(q_0,q).
\end{equation}
If a pole
is located at
\begin{equation}\label{eq312}
q_0=\omega_{T,L}+i\gamma_{T,L}
\end{equation}
then the imaginary shift of the pole $\gamma_{T,L}$ is
related to the
imaginary part of the self energy through
\begin{equation}\label{eq313}
\gamma_{T,L}=\mbox{Res}(\Delta_{T,L}) \mbox{Im}\Pi_{T,L},
\end{equation}
where $\mbox{Res}(\Delta)$ is the residue of the propagator
given by
\begin{equation}\label{eq314}
\mbox{Res}(\Delta_{T,L})^{-1}= \left.
\frac{\partial\Delta_{T,L}^{-1}}{\partial q_0}
\right|_{\omega_{T,L}} ,
\end{equation}
or, explicitly in terms of $\omega_{T,L}$
\begin{equation}
\label{eq314a}
\mbox{Res}(\Delta_T)^{-1} = -\omega_T + {q^2\over\omega_T}
+{3m_g^2\omega_T\over \omega_T^2-q^2}   ,
\end{equation}
\begin{equation}
\label{eq314b}
\mbox{Res}(\Delta_L)^{-1} = -\omega_L + {q^2\over\omega_L}
+{3m_g^2\over\omega_L}    .
\end{equation}
Thus we can write (\ref{eq38a}) as
\begin{equation}\label{eq315}
\Delta(Q)=\frac{Q^2-\mbox{Re} \Pi}
{(Q^2-\mbox{Re}\Pi)^2+
\mbox{Res}(\Delta)^{-2} \gamma^2}+
\frac{i \mbox{Res}(\Delta)^{-1} \gamma}
{(Q^2-\mbox{Re}\Pi)^2+
\mbox{Res}(\Delta)^{-2} \gamma^2},
\end{equation}
where we suppressed subscripts $T,L$, i. e.
 $\Delta,\Pi$ and $\gamma$ are either transverse or
longitudinal.
 This expression will be used
 to replace the mass-shell
$\delta$-function for thermal gluons.

 The imaginary part of the pole in (\ref{eq312})
gives the damping rate of the plasma
oscillations
\cite{pisarski,baier}
 In the following we  consider
 transverse gluons and we set
 $\gamma\equiv \gamma_T$.
 The damping rate
 is related to the so called gluon magnetic
 mass $m_{mag}$, or the inverse magnetic screening length
 at high temperature.
 Unfortunately, the exact relation between $\gamma$
 and $m_{mag}$
 is not known.
 A closed expression for the damping rate has been
 derived by Pisarski
 in the limit where
 $m_{mag} \gg \gamma$
\cite{pisarski}:
\begin{equation}\label{eq31}
\gamma = \frac{g^2 N T}{8 \pi}
\left[ \ln \left(\frac{m_g^2}{m_{mag}^2 +2m_{mag} \gamma}\right)
+1.1 \right] ,
\end{equation}
where the thermal gluon mass is given by
\begin{equation}
\label{eq32}
m_g^2=(N_c + \frac{N_f}{2})\frac{g^2 T^2}{9}.
\end{equation}
The magnetic mass at high temperature is of the form
\begin{equation}\label{eq33}
m_{mag}=c_N g^2 T
\end{equation}
where $c_N$ is a number depending on the gauge group and
cannot be calculated by a perturbation expansion.
Lattice estimates \cite{billoire,degrand} for SU(2)
\begin{equation}
\label{eq34}
c_2=0.27\pm0.03
\end{equation}
have been confirmed by recent semiclassical calculations
\cite{biro3}.
So far, no reliable estimate exists for SU(3)
\cite{grossman}.
The best one can do is extrapolate the SU(2) value
by \cite{danielewicz}
\begin{equation}\label{eq35}
c_3=\frac{3}{2} c_2
\end{equation}

Expanding the log in (\ref{eq31}) in powers of
$\gamma/m_{mag}$ and retaining only the leading terms
one finds
\begin{equation}\label{eq36}
\gamma = (1+\eta)^{-1}
 \frac{g^2 N T}{8 \pi}
\left[ \ln \left(\frac{m_g^2}{m_{mag}^2}\right)
+1.09681... \right] .
\end{equation}
where $\eta=0$ if we keep the leading log term only.
If the next-to-leading term is included then
\begin{equation}\label{eq37}
\eta = \frac{N}{4\pi c_N}.
\end{equation}
To check the consistency of Pisarski's approximation we plot
in Fig \ref{gamma}
the damping rate
for both values of $\eta$ as well as
the damping rate used in
\cite{altherr1,altherr2} and compare with $m_{mag}$ .
One can observe the poor validity of the approximation
for small values of $g$.
The approximation is well justified if one uses
the expression (\ref{eq36}) in the range
$1<g<2.5$.

\section{Production rates}
Consider a quark-gluon plasma in which the gluons and
the light quarks ($u,d$) are in thermal and chemical
equilibrium. The strange quarks too are in thermal
equilibrium
but away from chemical equilibrium
having very large and negative
chemical potential
$\mu\equiv\mu_s=\mu_{\bar{s}}$.
 The chemical reactions
\begin{equation}\label{eq20a}
q+\bar{q}\rightarrow s+\bar{s},
\end{equation}
\begin{equation}\label{eq20b}
g+g\rightarrow s+\bar{s},
\end{equation}
\begin{equation}\label{eq20c}
g\rightarrow s+\bar{s} ,
\end{equation}
will then take place until chemical equilibrium is reached.
The total production rate due to (\ref{eq20a}-\ref{eq20c}),
including the reversed processes, is given by
\cite{matsui}
\begin{equation}\label{eq20d}
\delta R=(1-{\rm e}^{2\beta \mu})(
R_{q\bar{q}\rightarrow s\bar{s}} +
R_{gg\rightarrow s\bar{s}} +
R_{g\rightarrow s\bar{s}} )
\end{equation}
where
\begin{eqnarray}
R_{q\bar{q}\rightarrow s\bar{s}} &=&
\int {d^3p_q\over (2\pi)^32E_q}
{d^3p_{\bar{q}}\over (2\pi)^32E_{\bar{q}}}
{d^3p_s\over (2\pi)^32E_s}
{d^3p_{\bar{s}}\over (2\pi)^3 2E_{\bar{s}}}
(2\pi)^4 \delta(P_q+P_{\bar{q}}-P_s-P_{\bar{s}})
 \nonumber \\
 & &
 \times
  f_{FD}(E_q)
 f_{FD}(E_{\bar{q}})(1-f_{FD}(E_s))(1-f_{FD}(E_{\bar{s}}))
\sum |M(q\bar{q}\rightarrow s\bar{s})|^2,
\label{eq21a}
\end{eqnarray}
\begin{eqnarray}
R_{gg\rightarrow s\bar{s}} &=& \half
\int {d^3p_1\over (2\pi)^32E_1}{d^3p_2\over (2\pi)^32E_2}
{d^3p_s\over (2\pi)^32E_s}
{d^3p_{\bar{s}}\over (2\pi)^3 2E_{\bar{s}}}
(2\pi)^4 \delta(P_1+P_2-P_s-P_{\bar{s}})
 \nonumber \\
 & &
 \times
 f_{BE}(E_1)
 f_{BE}(E_2)(1-f_{FD}(E_s))(1-f_{FD}(E_{\bar{s}}))
\sum |M(gg\rightarrow s\bar{s})|^2
\label{eq21b}
\end{eqnarray}
and
\begin{eqnarray}
R_{g\rightarrow s\bar{s}} &=&
\int {d^3q\over (2\pi)^32E_g}{d^3p_s\over (2\pi)^32E_s}
{d^3p_{\bar{s}}\over (2\pi)^3 2E_{\bar{s}}}
(2\pi)^4 \delta(Q-P_s-P_{\bar{s}})
 \nonumber \\
 & &
 \times
 f_{BE}(E_g)(1-f_{FD}(E_s))(1-f_{FD}(E_{\bar{s}}))
\sum |M(g\rightarrow s\bar{s})|^2 .
\label{eq21}
\end{eqnarray}
In our notation the four-momenta are
denoted by capitals $P, Q$ etc.
The summation in the above equations extends over
 all colors and polarizations
of the gluons and the final state quark-antiquark pair.
We have included the Pauli blocking factors although, as long as the
density of strange quarks
is well below one, Pauli blocking does not play a significant role.
Following
 Matsui, McLerran and Svetitsky
\cite{matsui} we can investigate the evolution process in terms
of the relaxation time determined near equilibrium.
Therefore we have to evaluate the rates
(\ref{eq21a}-\ref{eq21})
 at $\mu=0$, i. e. when quarks are in both
thermal and chemical equilibrium.
 Thermal field theory calculations
becomes fully legitimate in this way.
In particular, we can use the thermal quark mass given by
\cite{weldon2}
\begin{equation}\label{eq222}
m_s^2(T)=m_s^2(0)+\frac{g^2 T^2}{6}
\end{equation}
and the thermal gluon mass
(\ref{eq32}).

The processes of gluon fusion and quark-antiquark annihilation
have been discussed in ref
\cite{matsui}
and we shall
use their expressions for (\ref{eq21a},\ref{eq21b}).
The thermal gluon decay, also discussed by Altherr and Seibert
\cite{altherr1,altherr2}, can be calculated similarly. We first
replace
the integrations over $q$, $p_s$ and $p_{\bar{s}}$ by
\begin{eqnarray}
{d^3q\over 2E_g}&=&d^4\!Q\,\delta(Q^2-m_g^2)
\theta(q_0)
\label{eq22}  \\
{d^3p_s\over 2E_s}
{d^3p_{\bar{s}}\over 2E_{\bar{s}}}&=&
d^4\!P_s\,\delta(P_s^2-m_s^2)
\theta(p_s^0)\,
d^4\!P_{\bar{s}}\,\delta(P_{\bar{s}}^2-m_s^2)
\theta(p_{\bar{s}}^0)
\label{eq22a}
\end{eqnarray}
and change variables
\begin{eqnarray}\label{eq219}
Q' &=& P_s+P_{\bar{s}}
 \nonumber \\
P &=& \frac{1}{2}(
P_s-P_{\bar{s}}).
\end{eqnarray}
After trivially eliminating integrals over $d^4Q'$ and $d^3p$
we find
\begin{eqnarray}\label{eq220}
R_{g\rightarrow s\bar{s}} &=&
\frac{1}{4(2\pi)^4}\int
d^4\!Q\delta(Q^2-m_g^2)
\frac{1}{q}
  f_{BE}(q_0)
 \nonumber \\
 & &
 \times
\int dp_0
  (1-f_{FD}(\half{q_0}+p_0))(1-f_{FD}(\half{q_0}-p_o)
\sum |M(g\rightarrow s\bar{s})|^2
\end{eqnarray}
where the integration space is restricted by the following
kinematical constraints:
\begin{equation}\label{eq221}
q_0>2m_s ,
\;\;\;\;
0<q<(q_0^2-4m_s^2)^{1/2},
\;\;\;\;
p_0^2<\frac{q^2}{4}\left(1-\frac{4m_s^2}{Q^2}\right)
\end{equation}

It immediately follows that
$R_{g\rightarrow s\bar{s}}=0$
if $m_g<2m_s$.
 At the relevant temperatures the thermal
gluon mass is not high enough
 to allow for decay into a strange quark pair.
It is only because of its width that the gluon can decay.
To take this into account the $\delta$-function is replaced by a
function,
 similar to the Breit-Wigner resonance.
In the case of a narrow resonance the width of the resonance
 is related to the imaginary shift of the pole in the
propagator in the complex $q_0$ plane
\begin{equation}\label{eq23}
\frac{1}{q_0^2-(\sqrt{q^2+m^2}+i\gamma)^2}\approx
\frac{1}{Q^2-m^2}+
\frac{i2\sqrt{q^2+m^2}\,\gamma}{(Q^2-m^2)^2
+4(q^2+m^2)\gamma^2)}
\end{equation}
which in the limit $\gamma\rightarrow 0$ yields
the standard free particle propagator
\begin{equation}\label{eq24}
\frac{1}{Q^2-m^2-i\epsilon}=
       {\cal P}
\frac{1}{Q^2-m^2}+i\pi\delta(Q^2-m^2).
\end{equation}
Thus for a Breit-Wigner resonance  with width
$\Gamma=\gamma/2$ the mass-shell $\delta$-function
should be replaced by
\begin{equation}
\label{eq25}
\delta(Q^2-m^2)\rightarrow
\frac{1}{\pi}{\sqrt{q^2+m^2}\,\Gamma \over
 (Q^2-m^2)^2+(q^2+m^2)\Gamma^2}~~.
\end{equation}
This simple prescription cannot be directly applied to the
 case of thermal gluons because the location of the pole
is determined by complicated dispersion relations
(\ref{eq311})
for transverse (T) and longitudinal (L) gluons.
 Due to (\ref{eq315}),
 instead of (\ref{eq25}),
  we use
\begin{equation}\label{eq26}
\delta(Q^2-m_g^2)\rightarrow
\frac{1}{\pi}
\frac{\mbox{Res}(\Delta_{T,L})^{-1} \gamma_{T,L}}
{(Q^2-\mbox{Re}\Pi_{T,L})^2+
\mbox{Res}(\Delta_{T,L})^{-2} \gamma_{T,L}^2}.
\end{equation}

The matrix element is simply given by
\begin{equation}\label{eq211}
M(g\rightarrow s\bar{s})=g\epsilon_\mu(\zeta) \bar{u}(P_s)\gamma^\mu
\lambda_a v(P_{\bar{s}}) ,
\end{equation}
where $\epsilon_{\mu}(\zeta)$
   is the polarization vector of the decaying gluon
and $\lambda_a$ are the SU(3) matrices.

Summing over colors and all polarizations of the gluon leads to
\begin{eqnarray}
\sum_{a,\zeta}|M(g\rightarrow s\bar{s})|^2&=&
-4g^2\mbox{Tr}[(P_s+m_s)\gamma_\mu (P_{\bar{s}}-m_s)\gamma^{\mu}]
\nonumber \\
&=&16g^2(2m_s^2+Q^2) .
\label{eq212}
\end{eqnarray}

Since the frame of the quark-gluon plasma introduces
   a preferred direction,
it is furthermore necessary to distinguish between
the transverse and the longitudinal
 components of the gluons.
 If the sum is taken over transverse
 or longitudinal polarization only, we find
\begin{eqnarray}
\sum_{T}|M(g\rightarrow s\bar{s})|^2&=&
8g^2[4m_s^2+Q^2(1+4 \frac{p_0^2}{q^2})],
\label{eq213} \\
\sum_{L}|M(g\rightarrow s\bar{s})|^2&=&
8g^2Q^2(1-4 \frac{p_0^2}{q^2}).
\label{eq214}
\end{eqnarray}
By making use of
(\ref{eq26}) and (\ref{eq213},\ref{eq214}) we find
from (\ref{eq220})
\begin{eqnarray}\label{eq29}
R_{g\rightarrow s\bar{s}}^T &=&
\frac{g^2}{2\pi^4}\int_{2m_s}^{\infty} dq_0
  f_{BE}(q_0)
  \int_0^{\sqrt{q_0^2-4m^2_s}} dq\,q
  \int_{-\frac{q}{2}\sqrt{1-\frac{4m^2_s}{Q^2}}}^{
  \frac{q}{2}\sqrt{1-\frac{4m^2_s}{Q^2}}}
   dp_0
 \nonumber \\
 & &
 \times
  (1-f_{FD}(\half{q_0}+p_0))(1-f_{FD}(\half{q_0}-p_0))
 \nonumber \\
 & &
 \times
\frac{\mbox{Res}(\Delta_T)^{-1} \gamma_T}
{(Q^2-\mbox{Re}\Pi_T)^2+
\mbox{Res}(\Delta_T)^{-2} \gamma_T^2}
[4m_s^2+Q^2(1+4 \frac{p_0^2}{q^2})],
\end{eqnarray}
and a similar expression for
$R_{g\rightarrow s\bar{s}}^L $ .
The production rate due to the gluon decay is given by the sum
\begin{equation}\label{eq210}
R_{g\rightarrow s\bar{s}}=
R_{g\rightarrow s\bar{s}}^T+
R_{g\rightarrow s\bar{s}}^L.
\end{equation}

If we neglect the Pauli blocking factors
 the integral over $p_0$
 can be done explicitly, leading to
\begin{eqnarray}
\label{eq215}
R_{g\rightarrow s\bar{s}}^T &=&
\frac{2g^2}{3\pi^4}\int_{2m_s}^{\infty} dq_0
  f_{BE}(q_0)
  \int_0^{\sqrt{q_0^2-4m^2_s}} dq\,q^2
  \sqrt{1-\frac{4m_s^2}{Q^2}}
  (Q^2+2m_s^2)
 \nonumber \\
 & &
 \times
\frac{\mbox{Res}(\Delta_T)^{-1} \gamma_T}
{(Q^2-\mbox{Re}\Pi_T)^2+
\mbox{Res}(\Delta_T)^{-2} \gamma_T^2}
\end{eqnarray}
and a similar expression for
$R_{g\rightarrow s\bar{s}}^L $.

We  use  the full high temperature
expressions for
Re$\Pi_{T,L}$
given by (\ref{eq33},\ref{eq34}) and numerically
solve the dispersion relations (\ref{eq311}) in order
to determine
$\mbox{Res}(\Delta_{T,L})$ from (\ref{eq314a},\ref{eq314b}).
%
%
%
% For our purpose it is sufficient to
%
%
%
%
%Approximate expressions
% for hard gluons \cite{altherr2}, i. e. in the limit
%$q\rightarrow q_0$ are given by
%\begin{eqnarray}
%\mbox{Re}\Pi_T\approx\frac{3}{2}m_g^2,
%&\;\;\;\;&
%\mbox{Re}\Pi_L\approx 0,
%\nonumber  \\
%\mbox{Res}(\Delta_{T})^{-1}\approx2\sqrt{q^2+\frac{3}{2}m_g^2},
%&\;\;\;\;&
%\mbox{Res}(\Delta_{L})^{-1}\approx 2 q.
%\label{eq27}
%\end{eqnarray}
%
%
%
%
%
%
%
The temperature dependent gluon mass is given by
(\ref{eq32})
and the damping rate
$\gamma\equiv\gamma_T\approx\gamma_L$
is estimated using
(\ref{eq36}).
\section{Results and Conclusion}
The rates for diferent processes are
depicted in Fig \ref{rate1}
and Fig \ref{rate2}.
Our numerical calculation of the thermal gluon
decay is done using equations
(\ref{eq29},\ref{eq210}) with
(\ref{eq39},\ref{eq310},\ref{eq314a},\ref{eq314b}) and
(\ref{eq36}).
 The rates for quark-antiquark annihilation
and  gluon fusion we calculate by making use of
equations (3.23-3.25) in ref \cite{matsui}.
We fix the QCD coupling constant  at
the value $g=2$ because the temperature during
the time evolution is almost constant and the
 runing coupling effect is negligable.
In Fig \ref{rate1} the
quark mass is kept fixed while
in Fig \ref{rate2} it
 varies with temperature according to
(\ref{eq222}).
In both cases we find that
the gluon fusion together
with the quark-antiquark annihilation,
dominates almost everywhere.
The gluon decay process is as large as
the gluon fusion
in the narrow region around
$m_s(0)/T= 1$
only if we choose the optimistic
\cite{altherr2}
parameterization of the damping rate.

It has been shown that the time dependence
of the strange-quark density
  can, to a great degree of accuracy,
  be described by the approximate evolution equation \cite{matsui}
\begin{equation}
\label{eq41}
n_s(t)=n_s^{eq}{\rm tanh}
\left(\frac{t}{2\tau}+{\rm const}\right),
\end{equation}
where the relaxation time is defined as
\begin{equation}\label{eq42}
\tau=\frac{1}{2\beta R}
\left.\frac{\partial n_s}{\partial \mu}\right|_{\epsilon}
\end{equation}
with
\begin{equation}\label{eq43}
 R=
R_{q\bar{q}\rightarrow s\bar{s}} +
R_{gg\rightarrow s\bar{s}} +
R_{g\rightarrow s\bar{s}}.
\end{equation}
 The derivative of $n_s$ with respect to $\mu$
at fixed energy density is given by
\begin{equation}\label{eq44}
\left.\frac{\partial n_s}{\partial \mu}\right|_{\epsilon}
=\frac{\partial n_s}{\partial \mu}
-\frac{\partial n_s}{\partial T}
\left(\frac{\partial \epsilon}{\partial T}\right)^{-1}
\frac{\partial \epsilon}{\partial \mu}
\end{equation}
where
\begin{equation}\label{eq45}
n_s=2 N_f\int \frac{d^3p}{(2\pi)^3}f_{FD}(E_s,\mu)
\end{equation}
and
\begin{equation}\label{eq46}
\epsilon =
\frac{(N^2-1)\pi^2 T^4}{15}
+4N_f N\int \frac{d^3p}{(2\pi)^3}E_q f_{FD}(E_q,0)
+4N\int \frac{d^3p}{(2\pi)^3}E_s f_{FD}(E_s,\mu).
\end{equation}
All the quantities in (\ref{eq42}) are to be evaluated at
$\mu=0$.
In Fig  \ref{relax}  we plot the relaxation time
for the saturation of the strange-quark density for the
massive quarks with
the zero temperature mass
$m_s(0)=0.2$ GeV along with the classical approximation.
In this approximation
Pauli blocking factors $(1\!-\!f_{FD})$ are eliminated
and the remaining Fermi-Dirac and Bose-Einstein distributions
 are replaced by the Boltzmann distribution.
 For comparison we also plot the relaxation time for the
 massless quarks.

We comment here on the various approximations made in the gluon decay
calculation. First of all, use was made of the Braaten-Pisarski
resummation scheme. This is strictly valid only when $gT<< T$ which
is clearly not the case here.
This is the case with most applications
of QCD at finite temperatures.
Secondly the magnetic mass has been
introduced although only very limited knowledge is available. It has
been used to calculate the damping rate of a thermal gluon inside a
plasma.
In comparison to the calculations of references
\cite{altherr1,altherr2}
 we keep the
standard form of the Breit-Wigner distribution.
The consequences of this is that the rate for high
masses is reduced, while for low masses it is enhanced.
We also avoid
a rather heuristic  assumption that the thermal quark
mass is generated by gluons only. Since our rates are defined
near equilibrium our thermal mass includes both thermal gluon
and thermal quark contribution.
Finally, a more
 accurate calculation of the gluon fusion process shows that
 the high mass approximation used
in \cite{altherr2} underestimates its contribution.

Our main point has been that even with the parameters chosen
in Ref.\cite{altherr1,altherr2}
we do not support the claim that the gluon
decay process is the dominant mechanism for strange quark production
inside a quark-gluon plasma. To the best of our knowledge, the gluon
fusion mechanism is
the leading process.

\newpage
{\large \bf Acknowledgments}

We acknowledge useful discussions with
R.~Baier and T.~Altherr. After completing this paper we learned with
great
regret of  the untimely passing away of
Tanguy Altherr.
One of us (I.D.) acknowledges financial support
from the European Community under number \\
CI1$^*$-CT91-0893(HSMU) and from the
KFA (J\"ulich).
\newpage
\noindent {\bf Figure Captions:}
\begin{enumerate}
\item\label{gamma}
The damping rate and the magnetic mass versus the coupling
constant.
\item\label{rate1}
The quark production rate for thermal gluon decay
for different damping rates
(short dashed, long dashed and dot-dashed lines
correspond  to the damping rates depicted in
Fig \ref{gamma})
compared to
the production rate for gluon fusion and quark
antiquark annihilation (solid line).
The mass $m_s$ is temperature independent.
\item\label{rate2}
Same as Fig \ref{rate1} with the thermal mass $m_s$ given by
(\ref{eq222}).
\item\label{relax}
Relaxation times for the density of massive
(solid line) and massless (long dashed line) quarks.
Corresponding
 relaxation times in the classical
approximation are plotted with
dashed and dotted lines respectively.
\end{enumerate}
\end{document}